# EMPIRICAL BAYES FORMULATION OF THE ELASTIC NET AND MIXED-NORM MODELS: APPLICATION TO THE EEG INVERSE PROBLEM.

Deirel Paz-Linares[1], Mayrim Vega-Hernández[1], Pedro A. Rojas-López[1], Pedro A. Valdés-Sosa[1,2] and Eduardo Martínez-Montes[1,3]

(1) Cuban Neuroscience Center
(2) University of Electronic Science and Technology of China
(3) Politecnico di Torino, Italy

## ABSTRACT

The estimation of EEG generating sources constitutes an Inverse Problem (IP) in Neuroscience. This is an ill-posed problem, due to the non-uniqueness of the solution, and many kinds of prior information have been used to constrain it. A combination of smoothness (L2 norm-based) and sparseness (L1 norm-based) constraints is a flexible approach that have been pursued by important examples such as the Elastic Net (ENET) and mixed-norm (MXN) models. The former is used to find solutions with a small number of smooth non-zero patches, while the latter imposes sparseness and smoothness simultaneously along different dimensions of the spatio-temporal matrix solutions. Both models have been addressed within the penalized regression approach, where the regularization parameters are selected heuristically, leading usually to non-optimal solutions. The existing Bayesian formulation of ENET allows hyperparameter learning, but using computationally intensive Monte Carlo/Expectation Maximization methods. In this work we attempt to solve the EEG IP using a Bayesian framework for models based on mixtures of L1/L2 norms penalization functions (Laplace/Normal priors) such as ENET and MXN. We propose a Sparse Bayesian Learning algorithm based on combining the Empirical Bayes and the iterative coordinate descent procedures to estimate both the parameters and hyperparameters. Using simple but realistic simulations we found that our methods are able to recover complicated source setups more accurately and with a more robust variable selection than the ENET and LASSO solutions using classical algorithms. We also solve the EEG IP using data coming from a visual attention experiment, finding more interpretable neurophysiological patterns with our methods, as compared with other known methods such as LORETA, ENET and LASSO FUSION using the classical regularization approach.

# 1. INTRODUCTION

Inverse Problems (IP) are common in biophysics, where usually a little amount of data is available as compared to the huge amount of parameters needed to model the system (Tarantola, 2005). In neuroscience the estimation of electrical sources inside the brain, which are measured through Electroencephalography (EEG), is a classic example of IP. Finding the equation that relates the electrical potential at the scalp ($\mathcal{V}$) and the Primary Current Density (PCD), created by the electrical activity of large masses of neurons, is called the Forward Problem (FP) of EEG. Under certain assumptions about the geometry and electrical conductivity of the tissues in the head, and the quasistatic approximation, the solution of the FP reduces to obtain the kernel (linear operator) $\mathcal{K}$ involved in the expression (Riera and Fuentes, 1998):

$$\mathcal{V}(\mathbf{r}_e, t) = \int \mathcal{K}(r_e, r) \cdot \mathcal{J}(r, t) d\mathbf{r}^3 \qquad [1\text{-}1]$$

Here $\mathcal{J}(r,t)$ represents the PCD in both the space of brain generators and time, and the kernel is called Electric Lead Field (ELF), which depends on the position of the measuring electrodes $r_e$. The EEG-IP is then mathematically defined as obtaining $\mathcal{J}$ given a measurement $\mathcal{V}$ and a known ELF. Thus, equation [1-1] becomes a Fredholm Integral Equation of type I, which is commonly discretized as a system of linear equations:

$$V = KJ + \varepsilon \qquad [1\text{-}2]$$

where the parameters ($J$) constitute a $S \times T$ matrix ($T$ is the number of time instants and $S$ is the number of spatial generators that represents the discretized PCD. $V$ and $\varepsilon$ are $N \times T$ matrices, representing the measured EEG and the measurement noise, in $N$ electrodes (generally no more than 128) distributed according to a standard system (Klem et al., 1999). Correspondingly, $K$ is an $N \times S$ matrix obtained from the ELF discretization.

To achieve an acceptable spatial resolution of the estimated PCD (known as the *inverse solution*) the brain region is discretized in an array of thousands of volume units called voxels. This means that $N \ll S$ making the system of linear equations [1-2] underdetermined, i.e. does not have a unique solution. Then, the EEG-IP is an ill-posed problem in the sense of Hadamard (Hadamard, 1923) and the only way to find a unique solution is to introduce additional or prior information on the parameters, usually in the form of anatomical, physical and/or mathematical constraints.

## 1.1 Classical inference

One way to address the estimation of the PCD in [1-2] is the regularization approach or classical framework (Tikhonov and Arsenin, 1977), closely related to Penalized Least Squares (PLS) methods, that are generally expressed as:

$$\hat{J} = argmin_J \{\|V - KJ\|_2^2 + \lambda P(\mathbb{L}J)\} \qquad [1\text{-}3]$$

The first term in the cost function in [1-3] is the residual of the model and the second summarizes some imposed constraints (or penalty terms). The regularization parameter $\lambda$ controls the relative weight of the penalty function $P$. The matrix $\mathbb{L}$ adds information about the correlation structure of the parameter matrix. In the context of EEG-IP, this can be the identity matrix (independent PCD in each voxel) or the Laplacian operator (discrete second derivative) for requiring spatial smoothness of the PCD.

Many algorithms derived from the regularization approach exhibit advantages in the convergence time for minimizing the cost function in [1-3], when $P$ is differentiable and convex. A well-known example is the *Ridge* method (Hoerl and Kennard, 1970), where the penalty function is the L2 norm of the parameters. Particular cases of this method have given rise to several well-established solutions of the EEG-IP, such as Minimum-Norm, Weighted Minimum-Norm and Low Resolution Electromagnetic Tomography (LORETA). Their advantages and disadvantages have being thoroughly studied (Pascual-Marqui, 1999).

Other methods based on convex and non-differentiable penalties have shown good results at reconstructing sparse PCD's, such as the Least Absolute Shrinkage Selection Operator (LASSO) (Tibshirani, 1996) and its version LASSO FUSION (Land and Friedman, 1996), which use the L1 norm as the penalty function. Another advantage of the regularization approach is its extensibility to include multiple constraints, also called Multiple Penalized Least Squares (MPLS) (Vega-Hernández et al., 2008). Particular examples are the Elastic Net (ENET) (Zou and Hastie, 2005) and fused LASSO (Tibshirani, 2005), and more recently the smooth LASSO (Hebiri, 2008). All these models can be computed using general modified Newton-Raphson algorithms, such as Local Quadratic Approximation (LQA) (Fan and Li, 2001) and Majorization-Minorization (MM) (Hunter et al., 2005), without a considerable loss of speed in computations. However, the estimated solutions using these algorithms become very sensitive to the set of constraints selected to regularize them. In addition, the heuristic selection of optimal regularization parameters is computationally expensive and usually leads to an inadequate balance of the multiple restrictions.

## 1.2 Bayesian Inference

An alternative to regularization is the Bayesian approach, which address the IP in terms of finding the posterior pdf of the parameters ($J$) using the Bayes rule:

$$p(J|V, \alpha, \beta) \propto p(V|J, \beta) p(J|\alpha) \qquad [1\text{-}4]$$

Where $p(V|J, \beta)$ is the likelihood, $p(J|\alpha)$ is the prior pdf of parameters, and $(\alpha, \beta)$ are hyperparameters. The prior plays a similar role as a penalty function in the classical framework. As a particular case, this formalism leads to a fully equivalent solution to the optimization problem in equation [1-3], when the likelihood is a Normal pdf (with mean $KJ$ and unit variance) and the prior pdf has the following exponential form (using $\alpha = \lambda$):

$$p(J|\alpha) = \frac{1}{Z} e^{-\alpha P(\mathbb{L}J)} \qquad [1\text{-}5]$$

An important advantage of the Bayesian framework is that in a second level of inference (MacKay, 2003) one can estimate the hyperparameters of interest, typically variances and/or precisions of likelihood and priors. One way to achieve this is by using the Empirical Bayes procedure, where the aim is to maximize the posterior pdf of hyperparameters:

$$(\hat{\alpha}, \hat{\beta}) = argmax_{(\alpha,\beta)} p(\alpha, \beta | V) \quad [1\text{-}6]$$

This approach has been widely developed in the last years, especially with sparse priors, which has been called Sparse Bayesian Learning. An important example is the Relevance Vector Machine (Tipping, 2001; Schmolck and Everson, 2007), consisting in estimating sparse parameter vectors (using a univariate Normal prior pdf) while learning the hyperparameters (prior precisions of parameters, with a Gamma prior pdf). Although theoretically Bayesian learning methods are more robust for computing the inverse solution, fast and efficient inference algorithms have been currently developed only for simple models (Faul and Tipping, 2003). Bayesian formulations of the ENET (Li and Lin, 2010) or multiple variants of LASSO (Kyung et al., 2010), involves the use of computationally intensive Expectation-Maximization (EM) and/or Monte Carlo (MC) algorithms.

### 1.3 L1/L2 norms models approach to the EEG Inverse Problem

Neuroscience studies have provided evidences that either spatial smoothness or sparseness of the EEG electrical generators are properties consistent with different brain states, even coexisting in some cases. Therefore, important efforts have been made to solve the EEG IP by using constraints/prior information based on the combination of L1 and L2 norms simultaneously. Particular examples go from algorithms performing smooth and sparse estimations in separate steps (Liu et al., 2005; Valdés-Sosa et al., 2009) to explicit combining models as the sum of L1 and L2 penalty functions while using iterative algorithms to solve it (Nagarajan et al., 2006; Tian and Li 2011). The MPLS methodology also allowed to test different models for combining smoothness and sparseness, such as variants of ENET (Vega-Hernández et al., 2008).

In a seemingly different approach, some authors have used the idea of group penalization for this combination, e.g. using a penalty based on the L1 norm of a vector whose elements are obtained as the L2 norms of other vectors. One example is the Focal Vector Field reconstruction (Haufe et al., 2008) where sparseness is imposed on the amplitude of the PCD but keeping smoothness in the 3 spatial components (x, y, z) that defines the direction of this vector magnitude. The penalization function is then the L1 norm of the vector formed by the L2 norms of the PCD vector in each voxel. This type of penalization was extended to the spatio-temporal context, consisting in the application of an L1 norm along the first dimension (space) of the PCD (parameter matrix), and an L2 norm along the second dimension (time) (Ou et al., 2009).

These penalties were recognized as particular cases of priors based on *mixed-norms* (MXN), recently popularized thanks to the development of proximal operators and gradient based algorithms, which provides a generic an efficient approach to all MXN models (Gramfort et al., 2012).

Despite the current methods address both spatial and spatio-temporal analysis and use very different algorithms, in all of them the degree of sparseness and/or smoothness depends strongly on the regularization parameters, which are usually defined ad hoc or by means of heuristic methods and information criteria not always leading to optimal values. In this work we propose a Bayesian formalism to address models based on mixtures of L1/L2 norms penalization functions (Laplace/Normal priors), where different degrees of sparseness/smoothness can be imposed along individual dimensions (space and time) of the parameter map. Although we develop the theory for particular ENET and MXN models, the whole methodology can be generally applied to other types of Laplace/Normal priors. The algorithm can be considered as a Sparse Bayesian Learning method since it allows to estimate the regularization (hyper-) parameters that control sparse/smooth properties and perform variable selection, while at the same time avoiding EM or MC computations.

### 2. METHODS

### 2.1 The Bayesian Elastic Net

The Bayesian formulation of the ENET can be interpreted as a Laplace/Normal prior pdf model[1], with associated hyperparameters $(\alpha_1, \alpha_2)$ that are inversely related to the variances of the Normal and Laplace pdfs:

$$p(J|\alpha_1, \alpha_2) = \frac{1}{Z} e^{-\sum_t (\alpha_{1,t} \|J_{\cdot,t}\|_2^2 + \alpha_{2,t} \|J_{\cdot,t}\|_1)} \quad [2\text{-}1]$$

Where $\alpha_{1,t}$ and $\alpha_{2,t}$ represents the $t$-th element of the hyperparameters vectors $(\alpha_1, \alpha_2)$, the notation $J_{\cdot,t}$ refers to the $t^{\text{th}}$ column of $J$, and $\mathcal{Z}$ is the normalization constant.

Since the squared L2 norm and the L1 norm are separable as a sum of terms depending on the individual components, this prior pdf introduces stochastic independence in the whole spatio-temporal map of the parameters, i.e. [2-1] can be factorized for all voxels and time points (indexed by $i$ and $t$, respectively) with corresponding normalization constants $\mathcal{Z}_{i,t}$:

$$p(J|\alpha_1, \alpha_2) = \prod_{t,i} \frac{1}{Z_{i,t}} e^{-\alpha_{1,t} J_{i,t}^2 - \alpha_{2,t} |J_{i,t}|} \quad [2\text{-}2]$$

This prior pdf can be expressed as a scale Gaussian mixture (Andrews and Mallows, 1974), with a mixing pdf corresponding to a Truncated Gamma on the new hyperparameter $\gamma$. This formulation has been shown before in the Bayesian literature with different parametrizations (Li and

---

[1] Note that this mixture of Laplacian and Gaussian distributions is not the Normal-Laplace distribution described in (Reed, 2006).

Lin, 2010; Kyung et al., 2010). In our case, we found convenient to reorganize the variances and mixing distribution as detailed in the following Lemma.

***Lemma 2.1.1*** *(Proof in Appendix A): Let $J_{i,t}$ be the random variable distributing with pdf $\frac{1}{Z_{i,t}}e^{-\alpha_{1,t}J_{i,t}^2}e^{-\alpha_{2,t}|J_{i,t}|}$, where $Z_{i,t}$ is a normalization constant. Then the following equality holds:*

$$\frac{1}{Z_{i,t}}e^{-\alpha_{1,t}J_{it}^2-\alpha_{2,t}|J_{i,t}|} =$$

$$\int_0^{+\infty} N(J_{i,t}|0,\Lambda_{i,t})TGa\left(\gamma_{i,t}\left|\frac{1}{2},1,(k_t,\infty)\right.\right)d\gamma_{i,t} \quad [2\text{-}3]$$

$$\Lambda_{i,t} = \frac{1}{2\alpha_{1,t}}\overline{\Lambda}_{i,t}, \quad \overline{\Lambda}_{i,t} = \left(1 - \frac{k_t}{\gamma_{i,t}}\right) \quad [2\text{-}4]$$

$$k_t = \alpha_{2,t}^2/4\alpha_{1,t} \quad [2\text{-}5]$$

*Where $TGa\left(\gamma_{i,t}\left|\frac{1}{2},1,(k_t,\infty)\right.\right)$ is the Truncated Gamma pdf, with lower truncation limit $k_t$.*

With this particular formulation of ENET prior in a Bayesian model we change the relevant hyperparameters from the old regularization parameters $(\alpha_1,\alpha_2)$ to $(\alpha_1,k)$. The first one acts as the scale of variances of $J_{i,t}$ (eq. [2-4]), and the second acts as the lower truncation limit of the pdf for $\gamma_{i,t}$. Although this makes more challenging the interpretation of the influence of the hyperparameters, it speeds and ensures the convergence of the Empirical Bayes algorithm described in section 2.3. Given that $k$ is now directly learned (without estimating $\alpha_2$ and independently of $\alpha_1$), it is necessary to impose a convenient gamma prior for this hyperparameter. For $\alpha_1$ we can just choose a non-informative prior pdf.

Applied to the EEG IP stated in [1-2], the new ENET Bayesian model can be analytically described by the following set of pdfs.

Likelihood:
$$V_{\cdot,t} \sim N(V_{\cdot,t}|KJ_{\cdot,t},\beta_t I) \quad [2\text{-}6]$$

Prior pdf of parameters:
$$J_{\cdot,t} \sim N\left(J_{\cdot,t}\left|0,diag(\Lambda_{\cdot,t})\right.\right) \quad [2\text{-}7]$$

Prior pdf of hyperparameters:
$$\gamma_{\cdot,t} \sim \prod_i TGa\left(\gamma_{i,t}\left|\frac{1}{2},1,(k_t,\infty)\right.\right) \quad [2\text{-}8]$$

$$\alpha_{1,t} \sim \text{non-informative pdf}, \log p(\alpha_{1,t}) = const \quad [2\text{-}9]$$

$$k_t \sim Ga(\tau+1,\upsilon), \tau > 0, \upsilon > 0 \quad [2\text{-}10]$$

Here $\beta_t$ plays the role of the noise variance and $\tau$ and $\upsilon$ are the scale and shape parameters of the Gamma prior.

In the original Bayesian ENET model the parameter's prior pdf [2-2] controls the degree of sparseness/smoothness through variable selection according to the values of $\alpha_1$ and $\alpha_2$ directly, so that when $\alpha_1 \ll \alpha_2$ we have that $p(J|\alpha_1,\alpha_2) \approx La(J|0,\alpha_2)$, which is a prior that promotes sparse solutions, and when $\alpha_2 \ll \alpha_1$ we have that $p(J|\alpha_1,\alpha_2) \approx N\left(J\left|0,\frac{1}{2\alpha_1}\right.\right)$, which promotes smoothness for not-too-large values of $\alpha_1$. In our formulation of the ENET model the variable selection is performed in a similar fashion in the hyperparameters level, but in this case when $k$ is large ($\alpha_1 \ll \alpha_2$ for finite $\alpha_1$) we have by [2-8] that $\gamma \approx k$, which implies by [2-4] that $\Lambda \to 0$, finally promoting sparseness; and when $k$ is small ($\alpha_2 \ll \alpha_1$), $\Lambda \to \frac{1}{2\alpha_1}$ by [2-4] so that in [2-7] we have that $p(J|\gamma,\alpha_1,k) \approx N\left(J\left|0,\frac{1}{2\alpha_1}\right.\right)$.

## 2.2 Hierarchization of the Bayesian mixed-norm

***Definition 2.2.1:*** *Let $\mathcal{R}^{S\times T}$ be the real matrix-space with $S \times T$ dimensions. We will denote mixed-norm (MXN) as the function $\|\cdot\|_{p,q}: \mathcal{R}^{S\times T} \to \mathcal{R}_+$:*

$$\|x\|_{p,q} = \left(\sum_{t=1}^T\left(\sum_{i=1}^S|x_{i,t}|^p\right)^{\frac{q}{p}}\right)^{\frac{1}{q}}, p,q \geq 1, x \in \mathcal{R}^{S\times T} \quad [2\text{-}11]$$

The function [2-11] becomes an $L_p$ norm if instead of $x$ we take its projection over the column subspace ($\mathcal{R}^S$) and an $L_q$ norm for the projection over the row subspace ($\mathcal{R}^T$). This also holds if the mixed-norm is reformulated switching the order of the sum operators and exponents in [2-11], although the value of the mixed-norm is generally different when the argument is a full matrix.

The regularization approach with mixed-norms, as a penalization function in [1-3], becomes a convex, non-differentiable and non-separable (along columns or rows of the matrix-space) optimization problem, which adds a great computational cost to the inference process (Beck and Teboulle, 2009; Gramfort et al., 2012). Several particular cases lead to different practical problems. For example, in (Gramfort et al., 2012) the mixed-norm $\|J^T\|_{2,1}$ of the spatio-temporal PCD is used to propose a model equivalent to group Lasso (Kowalski and Torrésani, 2009), taking the L1 norm of the L2 norms of all rows, leading to spatially sparse solutions that are smooth in time. This model can be computed using efficient algorithms (e.g. FISTA, Gramfort et al., 2012), but it cannot be directly tackled with the Bayesian formalism. With the same goal of spatial sparseness and temporal smoothness, the Elitist Lasso model was proposed with a penalty based on the mixed-norm:

$$P(J) = \|J\|_{1,2}^2 \quad [2\text{-}12]$$

This model has been used in the context of time-frequency analysis of EEG signals and is solved using the Block Coordinate Relaxation algorithm (Kowalski and Torrésani, 2009).

Here, we propose a Bayesian formulation of this mixed-norm penalty which can be represented with the following multivariate pdf:

$$p(J) = \frac{1}{Z}e^{-\alpha\|J\|_{1,2}^2} \quad [2\text{-}13]$$

where $\alpha$ is a hyperparameter that controls simultaneously the level of spatial sparseness and temporal smoothness. Given that [2-12] admits the decomposition $P(J) = \sum_t \|J_{\cdot,t}\|_1^2$, (where $J_{\cdot,t}$ denotes the $t$-th column of $J$) this prior can be easily written for each time point separately. However, the penalization over one spatial component $J_{i,t}$ ($1 \leq i \leq S$) cannot be separated from the remaining components $J_{k,t}$ $(k \neq i)$. We then propose to rearrange $\|J_{\cdot,t}\|_1^2$ in the form $(\delta_{i,t} + |J_{i,t}|)^2$, for any index $i$, where $\delta_{i,t} = \sum_{k \neq i} |J_{k,t}|$ is a magnitude carrying the dependence with the other components of the vector parameters. This allows us to derive a hierarchical Bayesian model for this formulation based on the theory of Markov Random Fields.

***Definition 2.2.2****: Let $x$ be a random vector or random matrix, which has joint pdf of the form $p(x) = (1/Z)e^{-\mathcal{P}(x)}$, where $\mathcal{P}(x)$ can be decomposed as $\sum_{(i,k) \in \mathcal{I}} \mathcal{P}_{ik}(x_i, x_k)$; $\mathcal{I}$ is a set of pairs of index and $Z$ is a normalization constant. Then we say that $x$ is a 'Pair-wise Markov Random Field' (pMRF), and we define the functions $\{\mathcal{P}_{ik}(x_i, x_k)\}$ as potentials.*

***Definition 2.2.3****: Let $x$ be a random vector or random matrix. Let $x_{\mathcal{H}}$ denote a subset of elements of $x$, which has conditional joint pdf of the form $p(x_{\mathcal{H}}|x_{\mathcal{H}^c}) = (1/Z)e^{-\alpha \mathcal{P}(x)}$, where $x_{\mathcal{H}^c}$ is the complement of $x_{\mathcal{H}}$ in $x$, $\mathcal{P}(x)$ can be decomposed as $\sum_{(i,k) \in \mathcal{I}} \mathcal{P}_{ik}(x_i, x_k)$, $\mathcal{I} \subseteq \{(i,k): x_i \vee x_k \in x_{\mathcal{H}}\}$, and $Z$ is a normalization constant. Then we say that $(x_{\mathcal{H}}, x_{\mathcal{H}^c})$ is a 'Pair-wise Conditional Markov Random Field' (pCMRF). (Note that any couple of sets $(x_{\mathcal{H}}, x_{\mathcal{H}^c})$ from a pMRF constitutes a pCMRF; see the proof of Lemma 2.2.4 a) in Appendix B).*

In our case, the random vector consisting in a column of the spatio-temporal matrix $J$ constitutes a pMRF, since its prior pdf can be written as:

$$p(J_{\cdot,t}|\alpha) = \frac{1}{Z} e^{-\alpha \|J_{\cdot,t}\|_1^2} \qquad [2\text{-}14]$$

Where the exponent can be decomposed as:
$$\|J_{\cdot,t}\|_1^2 = \sum_i J_{i,t}^2 + \sum_{i>k} 2|J_{i,t}||J_{k,t}| \qquad [2\text{-}15]$$

Hence the potentials of the Definition 2.2.2 are proportional to:
$$P_{ik} = \begin{cases} J_{i,t}^2, & i = k \\ 2|J_{i,t}||J_{k,t}|, & i \neq k \end{cases} \qquad [2\text{-}16]$$

pMRFs underlies many models of the Statistical Physics approach (Kindermann and Snell, 1980) and also learning algorithms (Murphy, 2012). In the case of the model described by [2-14], given its particular properties in [2-15] and [2-16], we can reformulate the prior pdf in a hierarchical model at the parameters level. For this we need to use some relevant properties posed in the following Lemma.

***Lemma 2.2.4****: The following properties can be verified for the conditional probabilities in the pCMRF associated to the mixed-norm model of [2-14]. The proofs can be found in Appendix B.*

a) $p(J_{i,t}|J_{i^c,t}, \alpha) = \frac{1}{Z_i} e^{-\alpha J_{i,t}^2} e^{-2\alpha \delta_{i,t}|J_{i,t}|}$ [2-17]

Where $\delta_{i,t} = \sum_{k \neq i} |J_{k,t}|$ and $Z_i$ is a normalization constant.

b) $p(\delta_{\cdot,t}) = \frac{1}{\overline{Z}} e^{-\alpha \|W^{-1}\delta_{\cdot,t}\|_1^2} I_{\mathcal{R}_+^S}(\delta_{\cdot,t})$ [2-18]

Where $I_{\mathcal{R}_+^S}(\delta_{\cdot,t})$ is the indicator function of the region $\mathcal{R}_+^S = \{x \in \mathcal{R}^S : x_i \geq 0, i = \overline{1,S}\}$, the matrix $W$ can be expressed as the subtraction of the identity matrix from a ones-matrix: $W = 1_{S \times S} - \mathbb{I}_{S \times S}$, and $\overline{Z}$ is a normalization constant.

c) $p(J_{i,t}|\alpha) = \int p(J_{\cdot,t}|\alpha) dJ_{i^c,t} = \int p(J_{i,t}|J_{i^c,t}, \alpha) p(\delta_{\cdot,t}|\alpha) dJ_{i^c,t} d\delta_{\cdot,t}$ [2-19]

Using the properties a) and b) we show that by taking the conditional pdf [2-17], instead of [2-14], a simpler model is achieved, separable in the parameters level. The cost of such transformation is the addition of new hyperparameters $\delta_{i,t}$, which inherit the multivariate behavior (correlation structure) of $J$ in the original model. Property c) validates the use of the new model proposed in a) and b), showing that the marginal pdf of the parameters in the original model is identical to the marginalized joint pdf of the new one.

Realizing that [2-17] represents a combination of Normal and Laplace priors for $J_{i,t}$, a further transformation of the Bayesian mixed-norm model is possible, combining the results in Lemma 2.1.1 and Lemma 2.2.4 and dealing with $\alpha$ and $2\alpha\delta_{i,t}$ as hyperparameters, similarly to $\alpha_1$ and $\alpha_2$ in the original ENET model. The resulting prior of parameters and hyperparameters are as follows:

Prior pdf of parameters:
$$J_{\cdot,t} \sim N\left(J_{\cdot,t}\Big|0, diag(\Lambda_{\cdot,t})\right) \qquad [2\text{-}20]$$

$$\Lambda_{i,t} = \frac{1}{2\alpha} \overline{\Lambda}_{i,t}, \qquad \overline{\Lambda}_{i,t} = \left(1 - \frac{\alpha \delta_{i,t}^2}{\gamma_{i,t}}\right) \qquad [2\text{-}21]$$

Prior pdfs of hyperparameters:
$$\gamma_{\cdot,t} \sim TGa\left(\gamma_{i,t}\Big|\frac{1}{2}, 1, (\alpha \delta_{i,t}^2, \infty)\right) \qquad [2\text{-}22]$$

$$\delta_{\cdot,t} \sim \frac{1}{\overline{Z}} e^{-\alpha \|W^{-1}\delta_{\cdot,t}\|_1^2} I_{\mathcal{R}_+^S}(\delta_{\cdot,t}) \qquad [2\text{-}23]$$

$$\alpha \sim \text{non-informative pdf}, \log p(\alpha) = const \qquad [2\text{-}24]$$

Similarly to ENET, the matrix $diag(\Lambda_{\cdot,t})$ can be interpreted as an effective prior variance of the parameters. However, here, the degree of sparseness in variable selection is twofold. On one hand, through the regularization parameter $\alpha$, which is unique for the whole spatio-temporal map and imposes the same degree of sparseness to each column (time point) of parameters. On the other hand, the hyperparameters $\delta_{i,t}$ controls sparseness locally, so that when $\delta_{i,t}$ is large we will have $\gamma_{i,t} \approx \alpha \delta_{i,t}^2$, by [2-22], and the effective prior variances will tend to zero $\Lambda_{i,t} \to 0$, by [2-21], promoting higher degree of sparseness independently of $\alpha$.

## 2.3 Empirical Bayes learning algorithm using the Elastic Net and mixed-norm models

The Elastic Net and mixed-norm models in the Bayesian formulation of the spatio-temporal EEG IP admits the following factorization:

$$p(V, J, \Theta) = \prod_t p(V_{\cdot,t}, J_{\cdot,t}, \Theta_t) \quad [2\text{-}25]$$

Where $\Theta$ represents the hyperparameters of both Elastic Net and mixed-norm: $\Theta_t = \begin{cases} \gamma_{\cdot,t}, \alpha_{1,t}, k_t, \beta_t & \text{ENET} \\ \gamma_{\cdot,t}, \delta_{\cdot,t}, \alpha, \beta_t & \text{MXN} \end{cases}$

Each factor in [2-25] can be decomposed in three factors containing the likelihood, parameters prior pdf, and the hyperparameters prior pdf:

$$p(V_{\cdot,t}, J_{\cdot,t}, \Theta_t) =$$
$$N(V_{\cdot,t}|KJ_{\cdot,t}, \beta_t I) N(J_{\cdot,t}|0, diag(\Lambda_{\cdot,t})) p(\Theta_t) \quad [2\text{-}26]$$

The following proposition allows to easily derive the maximum a posteriori estimate for the parameters $J$ from this joints posterior pdf.

***Proposition 2.3.1*** (Magnus and Neudecker, 2007): *The following identity holds:*

$$N(V_{\cdot,t}|KJ_{\cdot,t}, \beta_t I) N(J_{\cdot,t}|0, diag(\Lambda_{\cdot,t})) =$$
$$N(V_{\cdot,t}|K\mu_{\cdot,t}, \beta_t I) N(\mu_{\cdot,t}|0, diag(\Lambda_{\cdot,t})) \times$$
$$|2\pi\Sigma(t)|^{\frac{1}{2}} N(J_{\cdot,t}|\mu_{\cdot,t}, \Sigma(t)) \quad [2\text{-}27]$$

Where $\mu_{\cdot,t} = \frac{1}{\beta_t}\Sigma(t)K^T V_{\cdot,t}$ is the posterior mean of parameters *(maximum a posteriori estimate)* and the posterior covariance matrix is:

$$\Sigma(t) = \left(\frac{1}{\beta_t}K^T K + \left(diag(\Lambda_{\cdot,t})\right)^{-1}\right)^{-1} \quad [2\text{-}28]$$

Moreover, using this proposition we can obtain the posterior pdf of hyperparameters by integrating [2-25] over $J$, after substituting [2-27] and [2-26]:

$$p(\Theta|V) \propto p(V, \Theta) = \int p(V, J, \Theta) dJ$$
$$= \prod_t \Big\{ N(V_{\cdot,t}|K\mu_{\cdot,t}, \beta_t I) N(\mu_{\cdot,t}|0, diag(\Lambda_{\cdot,t})) \times$$
$$|2\pi\Sigma(t)|^{\frac{1}{2}} p(\Theta_t) \Big\} \quad [2\text{-}29]$$

The Empirical Bayes procedure consists in minimizing the auxiliary function $\mathcal{L} = -\log p(\Theta|V)$, which is equivalent to maximizing [2-29]:

$$\mathcal{L} = \sum_t \Big\{ \frac{1}{2}\log|\beta_t I| + \frac{1}{2\beta_t}\|V_{\cdot,t} - K\mu_{\cdot,t}\|_2^2$$
$$+ \frac{1}{2}\log|diag(\Lambda_{\cdot,t})| + \frac{1}{2}\mu_{\cdot,t}^T \left(diag(\Lambda_{\cdot,t})\right)^{-1} \mu_{\cdot,t}$$
$$+ \frac{1}{2}\log|\Sigma^{-1}(t)| - \log p(\Theta_t) \Big\} \quad [2\text{-}30]$$

Depending on the model choice (ENET or MXN), we will have different expressions for the effective prior variances, according to [2-4] and [2-21]. In both cases, the prior pdfs for the hyperparameters in equation [2-30] follow from the equations [2-8] to [2-10] and [2-22] to [2-24], respectively:

$$-\log p(\Theta_t) =$$
$$\sum_i \log \int_{k_t}^\infty Ga\left(x\Big|\frac{1}{2}, 1\right) dx - \sum_i \log Ga\left(\gamma_{i,t}\Big|\frac{1}{2}, 1\right)$$
$$- \log p(\alpha_{1,t}) - \log p(k_t) - \log p(\beta_t) \quad \text{ENET}$$
or
$$\sum_i \log \int_{\alpha\delta_{i,t}^2}^\infty Ga\left(x\Big|\frac{1}{2}, 1\right) dx - \sum_i \log Ga\left(\gamma_{i,t}\Big|\frac{1}{2}, 1\right)$$
$$+ \log \overline{Z} + \alpha\|W^{-1}\delta_{\cdot,t}\|_1^2 - \log p(\alpha) - \log p(\beta_t) \quad \text{MXN}$$

To minimize the non-convex function $\mathcal{L}$ we use an iterative algorithm based on the coordinate descent strategy (Tipping, 2001; Wipf and Nagarajan, 2009), regarding the arguments $\{\mu, \overline{\Lambda}, \alpha, \alpha_1, k, \beta\}$, where we replaced $\gamma$ by $\overline{\Lambda}$ in [2-30] using [2-4] and [2-21]. We use the matrix $\Sigma$, which is a non-linear function of the hyperparameters, for the estimation of parameters and hyperparameters in the next step. In the particular case of the mixed-norm we do not minimize $\mathcal{L}$ over the hyperparameter $\delta$, due to the non-differentiability, but instead we just update it by using its relationship with parameters given in Lemma 2.2.4 a). The following proposition shows update formulas based on the matrix derivative of $\mathcal{L}$ (Magnus and Neudecker, 2007) with respect to hyperparameters in each model.

***Proposition 2.3.2:*** *Update formulas for parameters and hyperparameters in the ENET and MXN models:*

| | ENET | |
|---|---|---|
| a) | $\hat{\mu}_{\cdot,t} = \frac{1}{\beta_t}\Sigma(t)K^T V_{\cdot,t}$ | [2-31] |
| b) | $\widehat{\overline{\Lambda}}_{i,t} = \hat{\eta}_{i,t}/(k_t + \hat{\eta}_{i,t})$ | [2-32] |
|  | $\hat{\eta}_{i,t} = -\frac{1}{4} + \sqrt{\frac{1}{16} + (\mu_{i,t}^2 + \Sigma_{i,i}(t))\alpha_{1,t}^2 k_t^2}$ | |
| c) | $\hat{\alpha}_{1,t} = \left(\frac{S}{2}\right)/\sum_i\left\{\frac{(\mu_{i,t}^2 + \Sigma_{i,i}(t))}{\overline{\Lambda}_{i,t}}\right\}$ | [2-33] |
|  | $\hat{k}_t = zero[F(k_t)]$ | [2-34] |
|  | $F(k_t) = \sum_i\left\{\frac{1}{1-\overline{\Lambda}_{i,t}}\right\} + v - \left(\tau - \frac{S}{2}\right)\frac{1}{k_t}$ | |
|  | $-S(\pi k_t)^{-\frac{1}{2}} e^{-k_t}/\int_{k_t}^\infty Ga\left(x\Big|\frac{1}{2}, 1\right) dx$ | [2-35] |
| d) | $\hat{\beta}_t = \|V_{\cdot,t} - K\mu_{\cdot,t}\|_2^2 / \left(N + \sum_{i=1}^S \left(\frac{2\alpha_{1,t}\Sigma_{i,i}(t)}{\overline{\Lambda}_{i,t}}\right) - S\right)$ | [2-36] |

| | MXN | |
|---|---|---|
| a) | $\hat{\mu}_{\cdot,t} = \frac{1}{\beta_t}\Sigma(t)K^T V_{\cdot,t}$ | [2-37] |
| b) | $\widehat{\overline{\Lambda}}_{i,t} = \hat{\eta}_{i,t}/(\alpha\delta_{i,t}^2 + \hat{\eta}_{i,t})$ | [2-38] |
|  | $\hat{\eta}_{i,t} = -\frac{1}{4} + \sqrt{\frac{1}{16} + (\mu_{i,t}^2 + \Sigma_{i,i}(t))\alpha\delta_{i,t}^2}$ | |
| c) | $\hat{\alpha} = zero[F(\alpha)]$ | [2-39] |
|  | $F(\alpha) = \sum_{i,t}\left\{\frac{(\mu_{i,t}^2 + \Sigma_{i,i}(t))}{\overline{\Lambda}_{i,t}} + \frac{\delta_{i,t}^2}{1-\overline{\Lambda}_{i,t}}\right\} + \sum_{t=1}^T \|W^{-1}\delta_{\cdot,t}\|_1^2$ | |
|  | $-\frac{ST}{2\alpha} - \sum_{i,t}\left\{\frac{(\pi\alpha\delta_{i,t}^2)^{-\frac{1}{2}} e^{-\alpha\delta_{i,t}^2}\delta_{i,t}^2}{\int_{\alpha\delta_{i,t}^2}^\infty Ga(x|\frac{1}{2},1)dx}\right\}$ | [2-40] |
|  | $\hat{\delta}_{i,t} = \sum_{k \neq i}|\hat{\mu}_{k,t}|$ | [2-41] |
| d) | $\hat{\beta}_t = \|V_{\cdot,t} - K\mu_{\cdot,t}\|_2^2 / \left(N + \sum_{i=1}^S \left(\frac{2\alpha\Sigma_{i,i}(t)}{\overline{\Lambda}_{i,t}}\right) - S\right)$ | [2-42] |

The high computational cost for obtaining $\Sigma(t)$ by means of the matrix inversion operation in [2-28], can be avoided by using the economical singular value decomposition (SVD) of the lead field $K = LDR^T$, and the Woodbury identity (Magnus and Neudecker, 2007), leading to:

$$\Sigma(t) = \beta_t \, diag(\Lambda_{.,t}) R \left( R^T diag(\Lambda_{.,t}) R + \beta_t D^{-2} \right)^{-1} D^{-2} R^T \quad [2\text{-}43]$$

The update formulas in Proposition 2.3.2 a), b) are consistent with the sparseness constraint in both the ENET and MXN models, since the elements of the effective prior variance matrix $\Lambda$ (or equivalently $\bar{\Lambda}$) select which elements of $\mu$ become zero. When $\Lambda_{i,t} \to 0$ the $i$-th row and $i$-th -column of the matrix $\Sigma(t)$ in [2-43] tend to zero vectors, from where $\hat{\mu}_{i,t} \to 0$. In the same way, if some parameters are very small in a previous iteration ($\hat{\mu}_{i,t} \approx 0$, $\Sigma_{i,i}(t) \approx 0$), they will lead to $\Lambda_{i,t} \to 0$ in the next iteration (equations [2-32] y [2-38]). In some algorithms, this property usually means that if one activation is set to zero (e.g. removed from the active set) in an iteration, it will not appear as part of the solution. In our case, however, we do not prune to zero the small coefficients. Therefore, although unlikely, a "zeroed" activation might be re-estimated in a future iteration and contribute to the solution.

The non-linear terms of the formulas [2-35] and [2-40] in Proposition 2.3.2 c) are obtained from the derivative of the normalization constants in [2-8] and [2-22]. These terms decrease strictly with respect to their arguments leading to smaller values of $F$ for higher values of $k_t$ and $\alpha$, which is equivalent in both cases to have more zero elements in $\bar{\Lambda}$. The measurements variance $\beta_t$ in formulas [2-36] and [2-42] is generally considered superfluous in the learning process, because it only acts as a scale factor for the parameters and usually decelerates the algorithm convergence (Babacan et al., 2010). In our case, we fix it to $\beta_t=1$, for all time points.

We also use fixed values for the parameters of the Gamma distribution [2-10] in the ENET model. In particular we chose $\tau=S$, which preserves the monotony of [2-35] (in the sense that only one zero of $F$ exists), and $\upsilon=\epsilon S$, where $\epsilon$ is such that $k_t$ has a flexible prior, with mean($k_t$)$\approx 1/\epsilon$ and variance($k_t$)$\approx 1/(\epsilon^2 S)$. Obviously the value of $k_t$ that optimizes $\mathcal{L}$ will depend on $(\tau, \upsilon)$, since these have some influence in the intercept of [2-35]. In order to keep an adequate balance we impose identical prior to $\alpha_1$, which allows the flexibility in our learning of different degrees of sparseness/smoothness. Although optimal values for $(\tau, \upsilon)$ might also be estimated within the Empirical Bayes (setting respective priors for them), we only consider here an exploratory study where they are fixed in the ENET model. We call our methods ENET-RVM and MXN-RVM due to their conceptual similarity with the model and algorithm of the original RVM (Tipping, 2001). The pseudo codes for these algorithms are given in appendix D.

## 3. RESULTS

### 3.1 Simulation study

To validate the theory we test the ability of the proposed methods for reconstructing the PCD in a low dimensional, although realistic, simulation of the EEG IP. In a ring of 736 possible generators in the cerebral cortex (coordinates were defined in the brain standard atlas of the Montreal Neurological Institute), we simulate 3 sources (called "patches" A, B and C), having different spatial distribution (degree of sparseness) and time courses. The ELF was computed for 31 electrodes disposed on the scalp around the ring of generators, using the three homogeneous spheres model (Riera and Fuentes, 1998). The electric potential (EEG) was computed as the product of the ELF and the simulated PCD, adding white noise for a peak-signal-to-noise ratio (SNR) of 42 db. These and other details are shown in Figure 1.

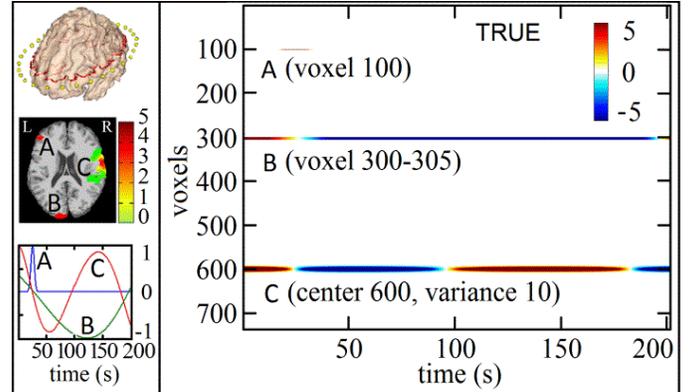

*Figure 1: Left, top to bottom: Generators space (red) and electrodes (yellow). Brain axial slice with spatial distribution of simulated sources, A (1 voxel), B (5 voxels), C (Gaussian). Time evolution of sources, A (narrow Gaussian), B and C (sinusoids). Right: Spatio-temporal color map representing simulated sources.*

#### 3.1.1 Bayesian Elastic Net solutions

The inverse solution was computed with the ENET-RVM algorithm in two versions: i) learning both hyperparameters and ii) fixing the hyperparameters to the combinations ($\alpha_1 = 1$, $\alpha_2 = 1$) and ($\alpha_1 = 1$, $\alpha_2 = 100$), which corresponds to the classical ENET. In the second scenario, the first combination tends to explain the data with solutions having the same balance between sparseness and smoothness, while the second with more sparse solutions. Figure 2 shows the simulated and estimated spatio-temporal maps for these scenarios.

With our learning strategy, the estimated values of the variance scale ($\alpha_1$) and the truncation coefficient ($k = \alpha_2^2/4\alpha_1$) varies in time as shown in Figure 3 (left). We also show in Figure 3 (right) the convergence of the hyperparameters' log-posterior $\mathcal{L}$ in a typical run of the ENET-RVM in both learned and fixed hyperparameters scenarios. Typically, the algorithm reached convergence around 10 iterations (1.07 min, in an Intel Core 2 Duo CPU, at 2.66GHz and 4GB of RAM memory).

We performed a comparison between solutions estimated with ENET-RVM and ENET-MM, where the parameters are estimated with the MM algorithm (Figure 4). Also, in ENET-MM the regularization parameters are modeled as $\alpha_1 = \lambda\mu$ and $\alpha_2 = \lambda(1 - \mu)$; where $\lambda$ is chosen such that it minimizes the Generalized Cross-Validation (GCV) function and $\mu$ is fixed to (0.1; 0.01; 0.001), which reinforces in the solution different degrees of sparseness with the ratio $\alpha_2/\alpha_1$ equal to 9; 99 and 999, respectively (Vega-Hernández et al., 2008).

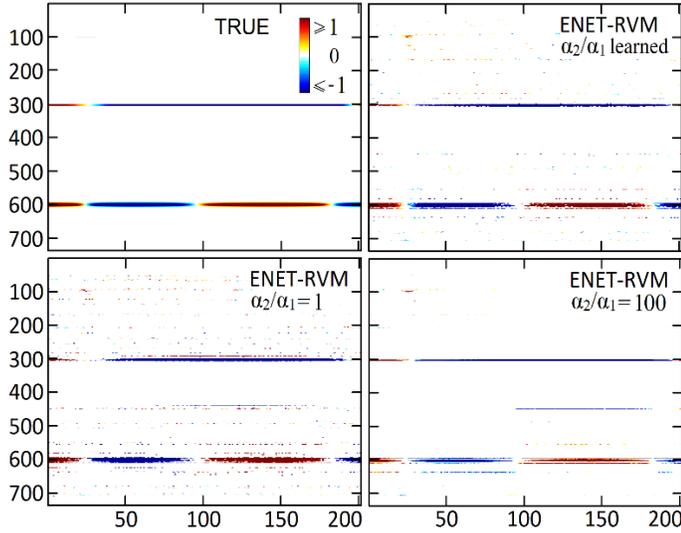

*Figure 2: Spatio-temporal maps of simulated (top-left) and estimated inverse solutions with the ENET-RVM algorithm, using different fixed values of the hyperparameters (bottom row) and with learning (top-right). Color maps are shown in a scale from -1 to 1 for an easier visualization.*

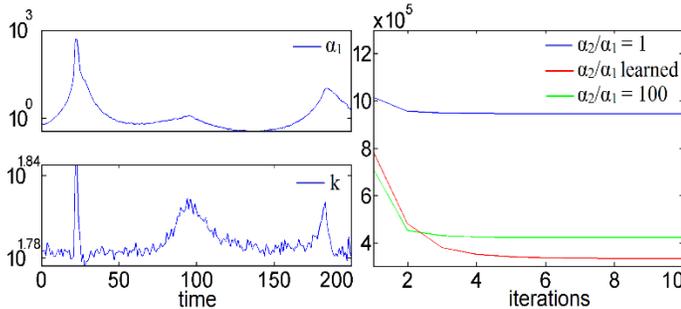

*Figure 3: Left: Temporal behavior of hyperparameters estimated with the ENET-RVM algorithm (logarithmic scale). Right: Convergence pattern of $\mathcal{L}$ in a typical run of the algorithm for the different strategies of selection of hyperparameters.*

For a quantitative evaluation of the estimated solutions, using both ENET-RVM and ENET-MM, we computed some quality measures based on the comparison with the simulated one (see Table I). The distance between both simulated and estimated spatio-temporal solutions (vectorized) was defined as 1 minus the correlation between them (1-corr), so that a small distance implies a high correlation (similarity measure). The sparseness (Sp.) of each solution was computed as the percentage of zero elements in the matrix, thus the Sp. of the simulation was 97.01%. In addition, we performed a Receiver Operating Characteristic (ROC) analysis and reported the accuracy or "area under the curve" (AUC), as well as the sensitivity (Sens.) and specificity (Spec.) when the solution is thresholded at 1% of its maximum value.

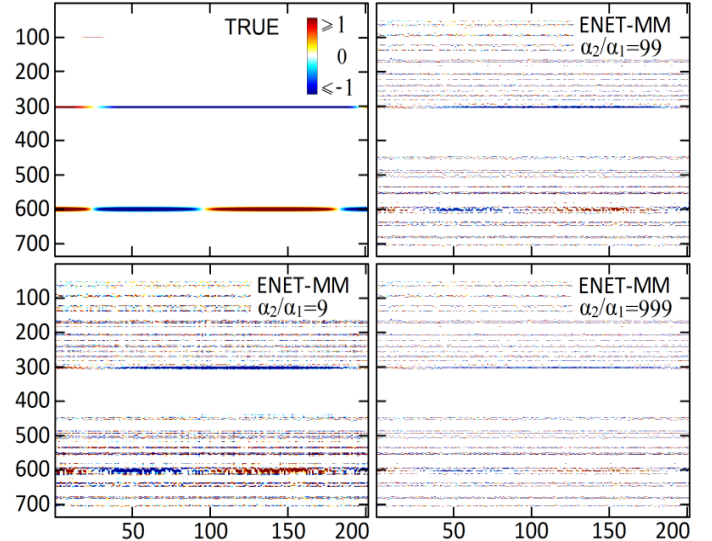

*Figure 4: Spatio-temporal maps of simulated (top-left) and estimated inverse solutions with the ENET-MM algorithm using different fixed values of the relative weight between the regularization parameters, where only one regularization parameter was selected as the one that minimizes the GCV function.*

*Table I: Quality measures of the ENET-RVM and ENET-MM solutions. The sparseness, sensitivity, specificity and area under ROC curve are expressed in percentage.*

|  | Solutions | 1-corr | Sp. | Sens. | Spec. | AUC |
|---|---|---|---|---|---|---|
| ENET-RVM | $\alpha_2/\alpha_1 = 1$ | 0.155 | 96.23 | 72.24 | 98.36 | 85.72 |
| | $\alpha_2/\alpha_1$ learned | 0.112 | 96.86 | 72.19 | 99.13 | 85.90 |
| | $\alpha_2/\alpha_1 = 100$ | 0.110 | 96.81 | 71.58 | 98.99 | 85.57 |
| ENET-MM | $\alpha_2/\alpha_1 = 9$ | 0.558 | 88.59 | 67.44 | 90.31 | 80.29 |
| | $\alpha_2/\alpha_1 = 99$ | 0.678 | 94.59 | 37.24 | 95.57 | 66.66 |
| | $\alpha_2/\alpha_1 = 999$ | 0.734 | 95.80 | 23.42 | 96.39 | 60.06 |

### 3.1.2 Bayesian mixed-norm solutions

For the MXN-RVM algorithm we also compared the inverse solutions obtained using fixed values of the hyperparameter ($\alpha = 1$ and $\alpha = 10$) with that of the learning scenario (see Figure 5). With our learning strategy the estimated value of the hyperparameter was $\alpha = 0.184$. Figure 6 shows that in a typical run of the MXN-RVM, the log-posterior $\mathcal{L}$ with learning and fixed values of hyperparameters converges around 10 iterations (also 1.07 min in the same computer as for ENET-RVM).
We explored the LASSO model solutions computed with the MM algorithm (LASSO-MM) (Vega-Hernández et al., 2008). Figure 7 shows the solutions for three different values of the regularization parameter, representative of a low ($\alpha = 0.002$), intermediate ($\alpha = 1.188$) and high ($\alpha = 572.599$) degrees of sparseness. Table II shows the quality measures for the different solutions obtained with MXN-RVM and LASSO-MM.

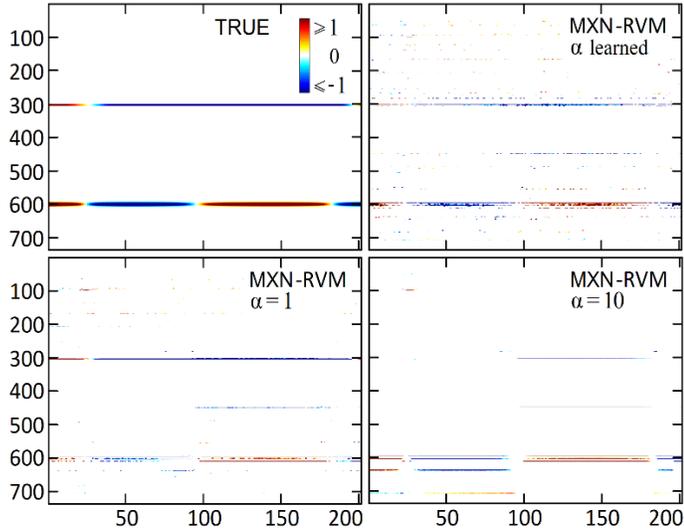

*Figure 5: Spatio-temporal maps of simulated (top-left) and estimated inverse solutions computed with the MXN-RVM algorithm, using different fixed values of the hyperparameters (bottom row) and with learning (top-right). Color maps are shown in a scale from -1 to 1 for an easier visualization.*

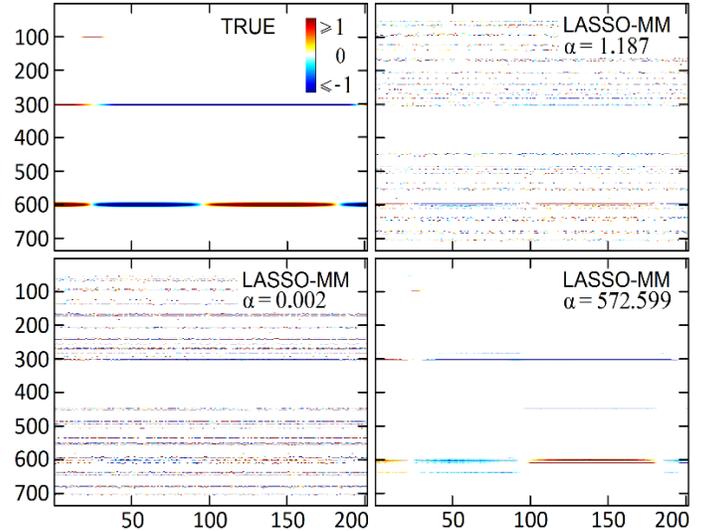

*Figure 7: Spatio-temporal maps of simulated (top-left) and estimated inverse solutions computed with the LASSO model using the MM algorithm (LASSO-MM) for different values of the regularization parameter. Color maps are shown in a scale from -1 to 1 for an easier visualization.*

## 3.2 Real data study of a visual attention experiment

EEG recordings using 30 electrodes (sampling frequency 128 Hz) were gathered from a healthy subject in a visual attention experiment, as explained in (Makeig et al., 1999). Briefly, a stimulus is presented and the subjects are requested to discriminate, by means of a physical action, between different geometric forms. The EEG recorded the brain activity for many repetitions of the stimulus, in 80 subjects, during a 3 s-long time window (1 s pre-stimulus and 2 s post-stimulus). Then, it is averaged over repetitions to cancel the background oscillatory activity. Figure 8 shows the 30 electrode's voltage time series (384 time instants) of the Grand Average Visual Evoked Potential, marking the stimulus onset (A) and the two time instants selected to compute the inverse solution with our methods. These corresponded to the global maximum negative peak (B: 281 ms post-stimulus) and global maximum positive peak (C: 430 ms post-stimulus). The median reaction times (i.e. when the subjects pressed a button) was about 350 ms. To compute the ELF we used an MNI standard brain, defining a grid of 3244 generators (voxels) within the volume of the gray matter, the brainstem and thalamus.

The ENET-RVM and MXN-RVM inverse solutions were compared with LORETA, one of the most used and well-studied solution. The results are visualized with the system Neuronic Tomographic Viewer (http://www.neuronicsa.com/), using the maximum intensity projection image (i.e. projecting brain activations to three orthogonal planes). Figures 9 and 10 show in color scale the PCD estimated with LORETA, ENET-RVM and MXN-RVM for the maximum negative and positive

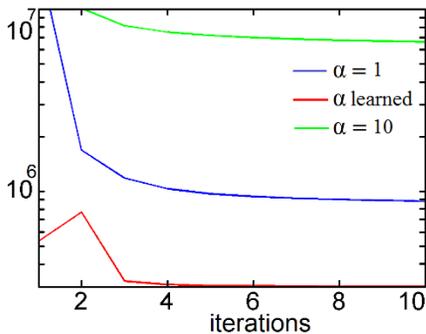

*Figure 6: Convergence pattern of $\mathcal{L}$ in a typical run of the MXN-RVM algorithm (logarithmic scale) for the different strategies of selection of the hyperparameter.*

*Table II: Quality measures of the MXN-RVM and LASSO-MM solutions. The sparseness, sensitivity and area under curve are expressed in percentage.*

|  | Solutions | 1-corr | Sp. | Sens. | Spec. | AUC |
|---|---|---|---|---|---|---|
| MXN-RVM | $\alpha = 1$ | 0.583 | 98.99 | 21.37 | 99.62 | 60.51 |
| MXN-RVM | $\alpha$ learned | 0.460 | 98.45 | 35.07 | 99.49 | 67.33 |
| MXN-RVM | $\alpha = 10$ | 0.562 | 98.93 | 21.52 | 99.56 | 60.56 |
| LASSO-MM | $\alpha = 0.002$ | 0.747 | 96.07 | 20.39 | 96.58 | 58.64 |
| LASSO-MM | $\alpha = 1.187$ | 0.59 | 96.93 | 21.91 | 97.51 | 59.87 |
| LASSO-MM | $\alpha = 572.599$ | 0.542 | 98.78 | 29.10 | 99.64 | 64.38 |

peak potentials (time points marked as B and C in Figure 8, respectively).

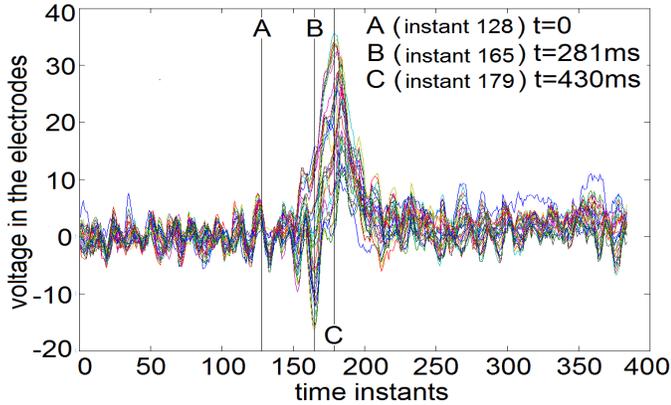

*Figure 8: Time series of the Visual Evoked Potential for all electrodes. A: Reference time instant (t=0) of stimulus onset. B: Global negative maximum potential (t = 281 ms). C: Global positive maximum potential (t = 430 ms).*

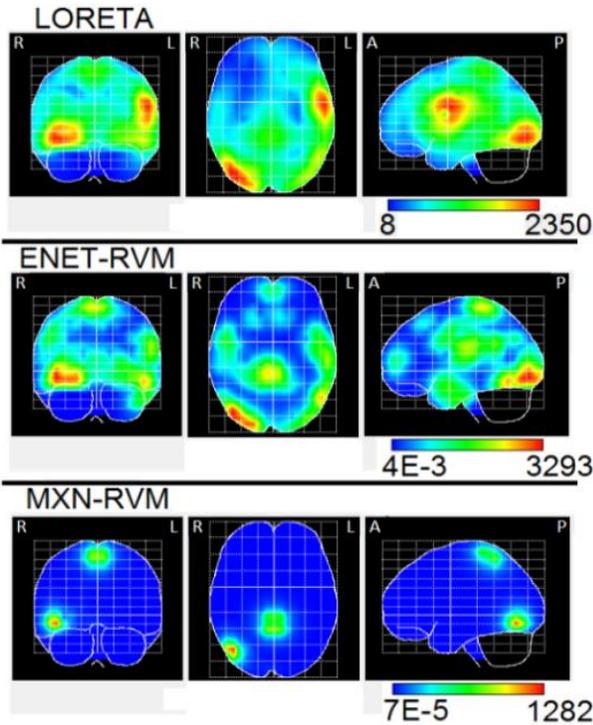

*Figure 9: Maximum intensity projection of the PCD estimated with LORETA, ENET-RVM and MXN-RVM, at the global negative maximum potential (time B in Figure 8). The three orthogonal planes are the coronal (left), axial (center) and sagittal (right) views. R, L, A, P stand for Right, Left, Anterior, Posterior, respectively.*

We also computed the solutions for the ENET-MM and for the LASSO FUSION model, also obtained using the MM algorithm. Figure 11 shows the ENET-MM solution with the same set of regularization parameters as in the simulation study, where $\lambda$ is chosen via GCV for different fixed values of $\mu$ (0.1; 0.01; 0.001). These values correspond to different ratios between regularization parameters $(\alpha_2/\alpha_1 = (1-\mu)/\mu)$ from 9 (low sparseness) to 999 (high sparseness).

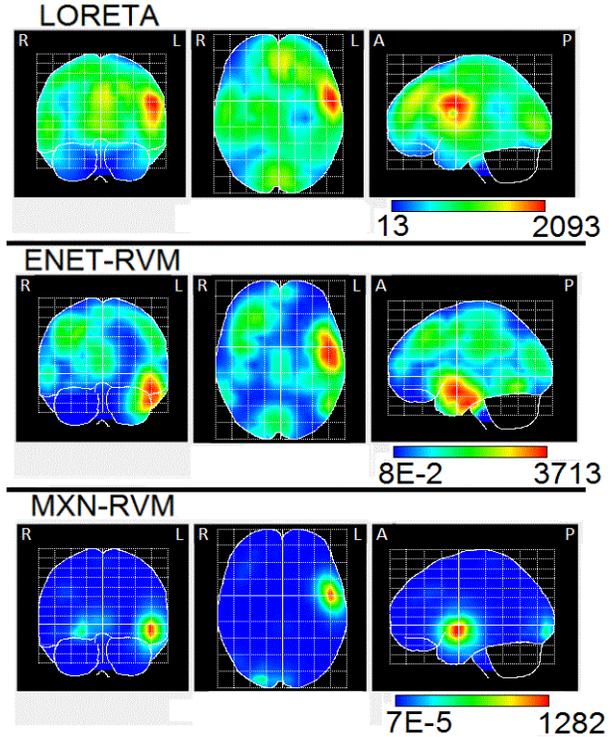

*Figure 10: Maximum intensity projection of the PCD estimated with LORETA, ENET-RVM and MXN-RVM, at the global positive maximum potential (time C in Figure 8). The three orthogonal planes are the coronal (left), axial (center) and sagittal (right) views. R, L, A, P stand for Right, Left, Anterior, Posterior, respectively.*

## 4. DISCUSSION

The Bayesian formulation of the ENET model proposed here is similar to (Li and Lin, 2010), and allows variable selection using the second level of inference for hyperparameters learning. Distinctively, we reformulate the hyperparameters as a scale factor ($\alpha_1$) of the parameters variances and the truncation coefficient ($k$) of the Truncated Gamma pdf that intervenes in variable selection. This allows the application of the Empirical Bayes procedure to estimate parameters and hyperparameters within an iterative coordinate descent algorithm, similar to the original RVM (Tipping, 2001; Wipf and Nagarajan, 2009). With this result we avoid the Double Shrinkage Problem of the classical ENET (Zou and Hastie, 2005; Vega-Hernández et al., 2008) and the use of computationally intensive MC/EM algorithms (Li and Lin, 2010; Kyung et al., 2010). Differently from the original RVM, here the hyperparameters ($\alpha_1$, $k$) controls the global degree of sparseness, while $\gamma$ acts in variable selection over voxels individually. A possible extension of our algorithm could be derived from assuming different values of the hyperparameters over individual voxels or groups of voxels (region of interest), to impose a variable degree of spatial sparseness.

In the same line, our Bayesian formulation of the MXN model, equivalent to the Elitist LASSO (Kowalski and Torrésani, 2009), pursues a similar constraint of spatial sparseness and temporal smoothness in the parameters matrix as previous works (Gramfort et al., 2012; Haufe et al., 2008; Ou et al., 2009), but achieving a separable model in the time dimension. This MXN model represents a Markov Random Field in the spatial dimension of parameters (Kindermann and Snell, 1980; Murphy, 2012), that can be reorganized into a Normal/Laplace hierarchical model at the parameters level. This allows to perform variable selection and learning of the hyperparameters by means of Empirical Bayes in MXN models, which has not been reported in the literature.

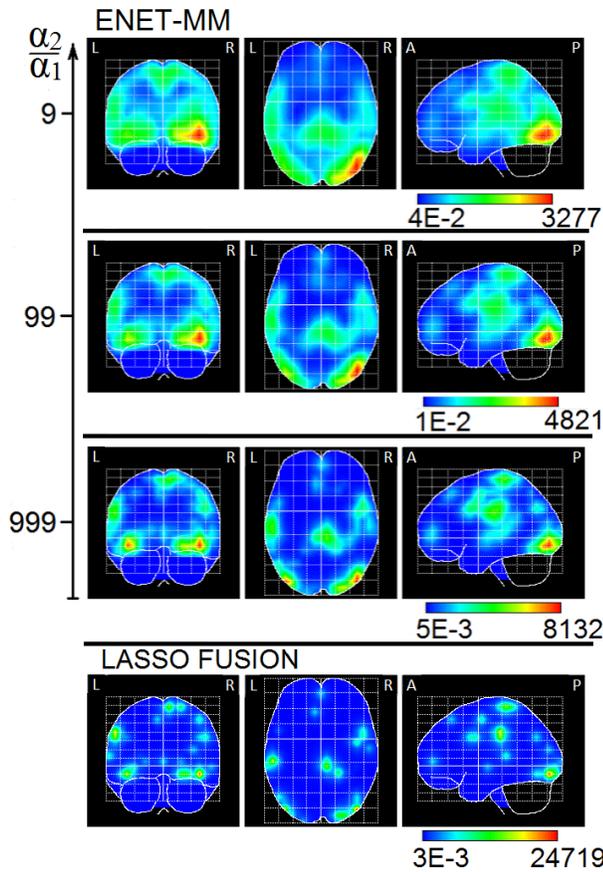

*Figure 11: Maximum intensity projection of the PCD estimated with ENET-MM (using different ratio between regularization parameters) and LASSO FUSION at the global negative maximum potential (time B in Figure 8). The three orthogonal planes are the coronal (left), axial (center) and sagittal (right) views. R, L, A, P stand for Right, Left, Anterior, Posterior, respectively.*

One interesting theoretical result in this work is that both ENET and MXN models can be expressed as particular cases of combined Normal/Laplace priors. Therefore, our algorithmic approach can be useful for many other models (including group LASSO, smooth LASSO, combinations of ENET and MXN, and others). Some of them have been tackled in the Bayesian approach, but only using MC/EM algorithms.

We tested the performance of the proposed methods in a simulated scenario where activations were not completely sparse in space nor completely smooth in time, thus challenging the assumptions of the models and exploring their capacity to adapt in non-ideal conditions. On one hand, ENET-RVM was able to recover spatially patch-wise smooth solutions, by accordingly tuning the degree of sparseness through the values of the hyperparameters, consistently with the theory (Figure 2). In the learning scenario, an intermediate degree of sparseness is achieved among the different simulated patches. The estimated values of hyperparameters were sensitive to the time-varying degree of sparseness, so that the values of the variances scale ($\alpha_1$) and the truncation coefficient ($k = \alpha_2^2/4\alpha_1$) behaved consistently with the theory (Figure 3, left), but with a low variation coefficient (mean=61.9, variance=2.8).

In the non-exhaustive analysis of the ENET-RVM algorithm presented here, we found a monotonous convergence pattern of the target function $\mathcal{L}$ (Figure 3, right), where the minimum value is achieved with learning of the hyperparameters. As shown in Table I, the solution in the learning scenario had a small distance to the simulated PCD, as compared to solutions using fixed values of the hyperparameters. It had the greatest AUC value, due to the high sensitivity and specificity, while the degree of sparseness is the most similar to the sparseness of the simulated solution (97.01%). Although the learning scenario seems to be promising, quantitative measures varied very little among the reported solutions from ENET-RVM. In the comparison with the solutions estimated with the known method ENET-MM (Hunter et al., 2005; Vega-Hernández et al., 2008; Sanchez–Bornot et al., 2008), we found that this one was able to recover the large patches (B and C) but introducing a larger amount of spurious or "ghost" activations than ENET-RVM (Figure 4). When imposing a high degree of sparseness (large $\alpha_2/\alpha_1$), solutions showed fewer ghost sources but the patch recovering deteriorated. In consequence, the quality measures in Table I showed that ENET-MM solutions had greater distances, lower sparseness and lower AUC than the three solutions obtained with the ENET-RVM algorithm.

On the other hand, the MXN-RVM was also consistent with the theory, being able to estimate sparse sources that are smooth during the time interval of the actual activation (Figure 5). The solution estimated with fixed hyperparameters exhibited more continuous temporal courses, but they were too sparse in space and presented more ghost sources than in the learning scenario. The latter performed generally better, since activations were spatially smooth, consistently with the learned value of the hyperparameter ($\alpha = 0.180$), which was smaller than the values fixed for computing the other solutions. We also found that MXN-RVM algorithm had a monotonous convergence pattern, and similarly to ENET-RVM the minimum value of the target function $\mathcal{L}$ was achieved in the learning scenario (Figure 6). As expected, solutions computed with LASSO-MM (Hunter et al., 2005; Vega-Hernández et al., 2008; Sanchez–Bornot et al., 2008) were highly sparse, although a large amount of ghost sources appeared for small values of the regularization parameter (Figure 7). The results in Table II suggest that MXN-RVM solution is more robust when sparseness assumptions hold, and in the learning scenario is able to recover highly sparse sources even when certain spatial support appears, with less ghost sources than the LASSO-MM solution.

If we compare the two proposed models, we will find that in general ENET-RVM performed better than MXN-RVM (compare the values in Tables I and II) since it could better recover the spatial extension of the simulated patches, while in MXN-RVM the quality measures were more sensitive to the values of the hyperparameter. We performed an additional exploratory study under different levels of noise (38 db and 34 db) for both models. The ENET-RVM solution under the combination $\alpha_2/\alpha_1 = 100$ was the most robust to noise, showing better values of the quality measures. The learning scenario, as in the study at 42 db, again adapted to the time-varying degree of sparseness and minimized $\mathcal{L}$ among other solutions. However, it tends to incorporate information from noise, reaching the lowest degree of sparseness and intermediate values of the distance and AUC in all cases. The algorithm MXN-RVM under higher noise levels was more robust than ENET-RVM. In the learning scenario, the quality measures were again better and $\mathcal{L}$ smaller than those for solutions with fixed values. Finally, computation times were similar in ENET-RVM and MXN-RVM, but the latter is more memory demanding because hyperparameter learning depends on all parameters of the spatio-temporal map.

In the analysis of real data, the maximum of the estimated solution with the three methods (LORETA, ENET-RVM and MXN-RVM) in the negative peak (time B in Figure 8), is located in the occipital area of the right hemisphere (Brodmann area 19), that corresponds to the primary visual cortex (Figure 9). As expected, LORETA is the most disperse solution, such that secondary activities appear not only in occipital areas but also in the superior temporal cortex. ENET-RVM showed similar solutions than LORETA, but less disperse, such that the secondary activities in temporal, frontal and centro-parietal areas can be distinguished among them. Contrary to LORETA, the secondary activation in the temporal cortex was not the most intense, but the centro-parietal one in the motor cortex (Brodmann area 4). The solution with the MXN-RVM method was the sparsest, showing only the main source in the right visual area and a secondary centro-parietal activation (motor cortex). Results from ENET-RVM and MXN-RVM are more consistent with the neurophysiology of visual attention where pre-motor activations in the occipital visual areas explain the processing of visual information, while secondary activations in the motor cortex are also expected related to the physical response (Hylliard and Anllo-Vento, 1998; Di Ruso et al., 2001).

A great similarity between ENET-RVM and ENET-MM solutions can be seen in the case when the ratio between regularization parameters in the latter is 99, while the hyperparameters are learned in the former (see Figures 9 and 11). Although preliminary, this suggests that the learning strategy might be an effective way to find optimal intermediate levels of sparseness in real scenarios, without the need of heavy statistical post-processing as in (Vega-Hernández et al., 2008). On the other hand, the MXN-RVM solution distinguishes from the very sparse but scattered LASSO FUSION solution (Figure 11). Encouragingly, this suggests that with MXN-RVM the degree of sparseness can be adapted to show a few small activated regions instead of many point sources, which is indeed more realistic in cognitive experiments.

The inverse solutions in the later positive peak (time C in figure 8) showed similar patterns as the computed in the negative peak with occipital and parieto-temporal activations (Figure 10). Again, this is in agreement with research on the neurological foundations of the visual evoked potentials, showing that late positive potentials can be related to processing of other cognitive aspects of visual information and attention, as well as post-motor responses, whose sources are likely to be found in the superior anterior temporal lobe (Makeig et al., 1999, 2004; Bonner and Price, 2013). However, in this case the three methods (LORETA, ENET-RVM and MXN-RVM) did not agree in the location of the main source. Yet very spread, LORETA solution showed the maximum activity in the superior temporal cortex (Brodmann area 43) and secondary activations in occipital, frontal and parietal areas, being difficult to decide which is relevantly separated from the main activation. Differently, ENET_RVM and MXN-RVM found the maximum activity in the anterior temporal lobe (Brodmann area 21). ENET_RVM showed again a similar solution as LORETA with a better trade-off between sparseness and smoothness, which would allow to identify different sources. The MXN-RVM solution is the sparsest again, clearly separating two PCD sources. The maximum activity was located in the superior gyrus of the anterior temporal lobe and the secondary activation in the occipital area. Again, sources were not scattered focal activations but well-localized smooth patches of activations. It seems that the MXN-RVM applied to just one time instant (i.e. without temporal information) does not behave simply like the LASSO model (Laplace prior), but it is able to find intermediate levels of spatial group sparseness. More comprehensive studies are needed to completely characterize the performance of this model with the proposed sparse Bayesian learning algorithm.

## 5. CONCLUSIONS

In this work we introduced a Bayesian formulation of models combining L1/L2 norm constraints, for solving the EEG-IP. In particular, we developed ENET and MXN models that have been previously addressed with the classical statistical framework, which presents practical limitations for selecting optimal values for one or more regularization parameters that are critical for correctly estimate solutions. We have used the

Empirical Bayes approach for deriving a Sparse Bayesian Learning algorithm that allows both the estimation of parameters and the learning of hyperparameters by means of iterative algorithms. Using simulations, we found that our methods ENET-RVM and MXN-RVM are able to recover complex sources more accurately and with a more robust variable selection than the classical ENET-MM and LASSO-MM algorithms. However, a limitation of our study is that comparisons were based on simple empirical criteria. Thus we think that future works should address more thorough validations with other simulated scenarios and using proper statistical analysis to assess the performance of the methods. In a real EEG study with a visual attention experiment, our methods localized electrical sources of early negative and late positive components of the Visual Evoked Potential that are more interpretable from the neurophysiological point of view, as compared with other known methods such as LORETA, ENET-MM and LASSO FUSION (also found using MM). Theoretically, the principles behind the proposed algorithms can be applied to other models based on combined Normal/Laplace priors and other types of inverse problems such as image recovery. Other possible extensions to deal with multidimensional data, such as time-frequency and space-time-frequency EEG tensors, should be carried out in the future.

# APPENDICES

## A. Proof of Lemma 2.1.1

The Normal/Laplace pdf in [2-3] can be rearranged as:
$$e^{-\alpha_1 J_{i,t}^2}e^{-\alpha_2|J_{i,t}|} = \frac{2}{\alpha_2}e^{-\alpha_1 J_{i,t}^2}La(J_{i,t}|\alpha_2)$$

Using the Gaussian scale mixture model for the Laplace pdf (Park and Casella, 2008):
$$La(J_{i,t}|\alpha_2) = \int_0^{+\infty} N(J_{i,t}|0, x_{i,t}) \frac{\alpha_2^2}{2} e^{-\frac{\alpha_2^2}{2}x_{i,t}} dx_{i,t}$$

We obtain:
$$e^{-\alpha_1 J_{i,t}^2}e^{-\alpha_2|J_{i,t}|} = \int_0^{+\infty} e^{-\alpha_1 J_{i,t}^2} \frac{\alpha_2}{\sqrt{2\pi x_{i,t}}} e^{-\frac{J_{i,t}^2}{2x_{i,t}}} e^{-\frac{\alpha_2^2}{2}x_{i,t}} dx_{i,t}$$

The term to the right can be rearranged by multiplying and dividing by $\sqrt{1 + 2\alpha_1 x_{i,t}}$ as:

$$\alpha_2 \int_0^{+\infty} \frac{1}{\sqrt{1+2\alpha_1 x_{i,t}}} \frac{1}{\sqrt{2\pi \frac{x_{i,t}}{1+2\alpha_1 x_{i,t}}}} e^{-\frac{J_{i,t}^2}{2\frac{x_{i,t}}{(1+2\alpha_1 x_{i,t})}}} e^{-\frac{\alpha_2^2}{2}x_{i,t}} dx_{i,t}$$

With the change of variables $\gamma_{i,t} = \frac{\alpha_2^2}{4\alpha_1}(1 + 2\alpha_1 x_{i,t})$ and $\Lambda_{i,t} = x_{i,t}/(1 + 2\alpha_1 x_{i,t}) = \frac{1}{2\alpha_1}\left(1 - \frac{\alpha_2^2}{4\alpha_1 \gamma_{i,t}}\right)$, we arrive at:

$$\sqrt{\frac{\alpha_2^2}{4\alpha_1}} e^{-\frac{\alpha_2^2}{4\alpha_1}} \int_{\frac{\alpha_2^2}{4\alpha_1}}^{+\infty} \frac{1}{\sqrt{2\pi \Lambda_{i,t}}} e^{-\frac{J_{i,t}^2}{2\Lambda_{i,t}}} \gamma_{i,t}^{-\frac{1}{2}} e^{-\gamma_{i,t}} d\gamma_{i,t} =$$

$$\int_{\frac{\alpha_2^2}{4\alpha_1}}^{+\infty} N(J_{i,t}|0, \Lambda_{i,t}) \sqrt{\frac{\alpha_2^2}{4\alpha_1}} e^{-\frac{\alpha_2^2}{4\alpha_1}} \gamma_{i,t}^{-\frac{1}{2}} e^{-\gamma_{i,t}} d\gamma_{i,t}$$

Then, using the definition for the Truncated Gamma (Gamma pdf truncated in the interval $\left(\frac{\alpha_2^2}{4\alpha_1}, \infty\right)$):

$$TGa\left(\gamma_{i,t}\left|\frac{1}{2}, 1, \left(\frac{\alpha_2^2}{4\alpha_1}, \infty\right)\right.\right) = \sqrt{\frac{\alpha_2^2}{4\alpha_1}} e^{-\frac{\alpha_2^2}{4\alpha_1}} \gamma_{i,t}^{-\frac{1}{2}} e^{-\gamma_{i,t}} I_{\left(\frac{\alpha_2^2}{4\alpha_1}, \infty\right)}(\gamma_{i,t})$$

We demonstrate that the Normal/Laplace pdf can be represented as the following scale mixture of Gaussians:
$$e^{-\alpha_1 J_{i,t}^2}e^{-\alpha_2|J_{i,t}|} =$$
$$\int_0^{+\infty} N(J_{i,t}|0, \Lambda_{i,t}) TGa\left(\gamma_{i,t}\left|\frac{1}{2}, 1, \left(\frac{\alpha_2^2}{4\alpha_1}, \infty\right)\right.\right) d\gamma_{i,t} \quad \blacksquare$$

## B. Proof of Lemma 2.2.4

a) Setting $x_\mathcal{H} = J_{i,t}$ and $x_{\mathcal{H}^c} = J_{i^c,t}$ into Definition 2.2.2, the conditional probability of one variable (node) regarding their complement in the pMRF can be expressed as:
$$p(J_{i,t}|J_{i^c,t}, \alpha) = p(J_{,t}|\alpha)/p(J_{i^c,t}|\alpha) \qquad [\text{B-1}]$$
Using [2-15] and [2-16], the conditional pdf in [2-14] can be decomposed as:
$$p(J_{,t}|\alpha) \propto e^{-\alpha P_{ii} - 2\alpha \sum_{k \neq i} P_{ik} - \alpha \sum_{l \neq i, k \neq i} P_{lk}} \qquad [\text{B-2}]$$
Which can be marginalized over $J_{i,t}$ to obtain:
$$p(J_{i^c,t}|\alpha) \propto e^{-\alpha \sum_{l \neq i, k \neq i} P_{lk}} \int e^{-\alpha P_{ii} - 2\alpha \sum_{k \neq i} P_{ik}} dJ_{i,t} \quad [\text{B-3}]$$
Substituting [B-2] and [B-3] in [B-1] we can find the potentials of Definition 2.2.3 for the duet $(J_{i,t}, J_{i^c,t})$:
$$p(J_{i,t}|J_{i^c,t}, \alpha) = \frac{e^{-\alpha P_{ii} - 2\alpha \sum_{k \neq i} P_{ik}}}{\int e^{-\alpha P_{ii} - 2\alpha \sum_{k \neq i} P_{ik}} dJ_{i,t}}$$
$$= \frac{1}{Z_i} e^{-\alpha J_{i,t}^2} e^{-\alpha \sum_{k \neq i} 2|J_{i,t}||J_{k,t}|}$$
Where $Z_i$ is a normalization constant. Then, using the auxiliary magnitude $\delta_{i,t} = \sum_{k \neq i} |J_{k,t}|$, we finally obtain:
$$p(J_{i,t}|J_{i^c,t}, \alpha) = \frac{1}{Z_i} e^{-\alpha J_{i,t}^2} e^{-2\alpha \delta_{i,t}|J_{i,t}|} \quad \blacksquare$$

b) The auxiliary matrix $(\delta)$ can be seen as new hyperparameters that are related to parameters in the form:
$\delta_{,t} = W|J_{,t}|$ ; with $W_{S \times S} = 1_{S \times S} - \mathbb{I}_{S \times S}$
Where $1_{S \times S}$ is an $S \times S$ matrix of ones and $\mathbb{I}_{S \times S}$ is the identity matrix. We can express the marginal pdf of $\delta_{,t}$ as:
$$p(\delta_{,t}|\alpha) = \int p(\delta_{,t}, J_{,t}|\alpha) dJ_{,t} = \int p(\delta_{,t}|J_{,t}, \alpha) p(J_{,t}|\alpha) dJ_{,t}$$
Since the conditional pdf $p(\delta_{,t}|J_{,t}, \alpha)$ can be represented by means of the Dirac delta function $\overline{\delta}$:
$$p(\delta_{,t}|J_{,t}, \alpha) = \overline{\delta}(\delta_{,t} - W|J_{,t}|) I_{\mathcal{R}_+^S}(\delta_{,t})$$
We can use this and [2-14] to obtain:
$$p(\delta_{,t}|\alpha) = \int \overline{\delta}(\delta_{,t} - W|J_{,t}|) I_{\mathcal{R}_+^S}(\delta_{,t}) \frac{1}{Z} e^{-\alpha \|J_{,t}\|_1^2} dJ_{,t}$$
Taking into consideration the symmetry of the argument with respect to the origin, and rearranging the Dirac delta function we obtain:

$p(\delta_{.,t}|\alpha) = \frac{1}{|W|\mathcal{Z}} I_{\mathcal{R}_+^S}(\delta_{.,t}) \int \bar{\delta}(|J_{.,t}| - W^{-1}\delta_{.,t})e^{-\alpha\|J_{.,t}\|_1^2} dJ_{.,t}$

$= \frac{1}{\bar{\mathcal{Z}}} e^{-\alpha\|W^{-1}\delta_{.,t}\|_1^2} I_{\mathcal{R}_+^S}(\delta_{.,t})$

where $\bar{\mathcal{Z}} = |W|\mathcal{Z}/2^S$ ∎

c) The joint pdf of parameters and new hyperparameters $\delta$ is:

$p(J_{.,t}, \delta_{.,t}|\alpha) = \prod_k p(J_{k,t}|J_{k^c,t}, \alpha) p(\delta_{.,t}|\alpha)$

Using final results from previous items a) and b), it becomes:

$p(J_{.,t}, \delta_{.,t}|\alpha) =$

$\prod_k \frac{1}{Z_k} e^{-\alpha J_{k,t}^2} e^{-2\alpha\delta_{k,t}|J_{k,t}|} \frac{1}{\bar{\mathcal{Z}}} e^{-\alpha\|W^{-1}\delta_{.,t}\|_1^2} I_{\mathcal{R}_+^S}(\delta_{.,t})$

Marginalizing over $J_{i^c,t}$:

$\int p(J_{.,t}, \delta_{.,t}|\alpha) dJ_{i^c,t} = \frac{1}{\bar{\mathcal{Z}}} e^{-\alpha\|W^{-1}\delta_{.,t}\|_1^2} I_{\mathcal{R}_+^S}(\delta_{.,t}) \times$

$\int \prod_k \left[\frac{1}{Z_k} e^{-\alpha J_{k,t}^2} e^{-2\alpha\delta_{k,t}|J_{k,t}|}\right] dJ_{i^c,t} =$

$= \frac{1}{\bar{\mathcal{Z}}} e^{-\alpha\|W^{-1}\delta_{.,t}\|_1^2} I_{\mathcal{R}_+^S}(\delta_{.,t}) \frac{1}{Z_i} e^{-\alpha J_{i,t}^2} e^{-2\alpha\delta_{i,t}|J_{i,t}|}$

And marginalizing over $\delta_{.,t}$ using the change of variable $\delta_{.,t} = W|J'_{.,t}|$, we obtain:

$\int p(J_{.,t}, \delta_{.,t}|\alpha) dJ_{i^c,t} d\delta_{.,t} =$

$\frac{|W|}{2^S \bar{\mathcal{Z}}} \int \frac{1}{Z_i} e^{-\alpha J_{i,t}^2 - 2\alpha(\sum_{k\neq i}|J'_{k,t}|)|J_{i,t}| - \alpha\|J'_{.,t}\|_1^2} dJ'_{.,t}$

Where we remember that $\bar{\mathcal{Z}} = |W|\mathcal{Z}/2^S$; insert the decomposition:

$\|J'_{.,t}\|_1^2 = J'^2_{i,t} + 2(\sum_{k\neq i}|J'_{k,t}|)|J'_{i,t}| + (\sum_{k\neq i}|J'_{k,t}|)^2$

And rearrange terms to arrive at:

$\int p(J_{.,t}, \delta_{.,t}|\alpha) dJ_{i^c,t} d\delta_{.,t} =$

$\frac{1}{\mathcal{Z}} \int \left\{ e^{-\alpha J_{i,t}^2 - 2\alpha(\sum_{k\neq i}|J'_{k,t}|)|J_{i,t}| - \alpha(\sum_{k\neq i}|J'_{k,t}|)^2} \times \right.$

$\left. \frac{1}{Z_i} \left( \int e^{-\alpha J'^2_{i,t} - 2\alpha(\sum_{k\neq i}|J'_{k,t}|)|J'_{i,t}|} dJ'_{i,t} \right) \right\} dJ'_{i^c,t}$

Now, realizing that $\int e^{-\alpha J'^2_{i,t} - 2\alpha(\sum_{k\neq i}|J'_{k,t}|)|J'_{i,t}|} dJ'_{i,t} = Z_i$ and changing back $J'_{i^c,t}$ to $J_{i^c,t}$, we can easily obtain:

$\int p(J_{.,t}, \delta_{.,t}|\alpha) dJ_{i^c,t} d\delta_{.,t} =$

$\int \frac{1}{\mathcal{Z}} e^{-\alpha\|J_{.,t}\|_1^2} dJ_{i^c,t} = \int p(J_{.,t}|\alpha) dJ_{i^c,t}$ ∎

## C. Derivation of updates equations in Proposition 2.3.2

a) Parameters. In both ENET and MXN we obtain:

$\frac{\partial \mathcal{L}}{\partial \mu_{.,t}} = \frac{\partial}{\partial \mu_{.,t}} \frac{1}{2} \mu_{.,t}^T \left(diag(\Lambda_{.,t})\right)^{-1} \mu_{.,t} + \frac{\partial}{\partial \mu_{.,t}} \frac{1}{2\beta_t} \|V_{.,t} - K\mu_{.,t}\|_2^2$

$= \left(diag(\Lambda_{.,t})\right)^{-1} \mu_{.,t} + \frac{1}{\beta_t} K^T K \mu_{.,t} - \frac{1}{\beta_t} K^T V_{.,t}$

$= -\frac{1}{\beta_t} K^T V_{.,t} + \left(\frac{1}{\beta_t} K^T K + \left(diag(\Lambda_{.,t})\right)^{-1}\right) \mu_{.,t}$

Using $\Sigma(t) = \left(\frac{1}{\beta_t} K^T K + \left(diag(\Lambda_{.,t})\right)^{-1}\right)^{-1}$ and equating to zero we obtain:

$\hat{\mu}_{.,t} = \frac{1}{\beta_t} \Sigma(t) K^T V_{.,t}$ ∎

b) Hyperparameters. For ENET we will have:

$\frac{\partial \mathcal{L}}{\partial \bar{\Lambda}_{i,t}} = \frac{\partial}{\partial \bar{\Lambda}_{i,t}} \frac{1}{2} log|\Sigma^{-1}(t)| + \frac{\partial}{\partial \bar{\Lambda}_{i,t}} \frac{1}{2} log|diag(\Lambda_{.,t})| +$

$+ \frac{\partial}{\partial \bar{\Lambda}_{i,t}} \frac{1}{2} \mu_{.,t}^T \left(diag(\Lambda_{.,t})\right)^{-1} \mu_{.,t} - \frac{\partial}{\partial \bar{\Lambda}_{i,t}} logGa\left(\gamma_{i,t}/\frac{1}{2}, 1\right)$

Deriving and reorganizing terms:

$\frac{\partial \mathcal{L}}{\partial \bar{\Lambda}_{i,t}} = -(\Sigma_{i,i}(t) + \mu_{i,t}^2) \frac{\alpha_{1,t}}{\bar{\Lambda}_{i,t}^2} + \frac{1}{2\bar{\Lambda}_{i,t}} + \frac{1}{2(1-\bar{\Lambda}_{i,t})} + \frac{k_t}{(1-\bar{\Lambda}_{i,t})^2}$

Equating to zero and using the change of variable $\bar{\Lambda}_{i,t} = \eta_{i,t}/(\eta_{i,t} + k_t)$ we obtain the equation:

$-(\Sigma_{i,i}(t) + \mu_{i,t}^2)\alpha_{1,t} k_t^2 + \frac{k_t \eta_{i,t}}{2} + k_t \eta_{i,t}^2 = 0$

Where the only positive root is:

$\hat{\eta}_{i,t} = -\frac{1}{2} + \sqrt{\frac{1}{4} + \left(\mu_{i,t}^2 + \Sigma_{i,i}(t)\right)\alpha_{1,t} k_t}$

So that if we define $\hat{\gamma}_{i,t} = \hat{\eta}_{i,t} + k_t$ we will have:

$\bar{\Lambda}_{i,t} = \hat{\eta}_{i,t}/\hat{\gamma}_{i,t}$

For the MXN model we follow the same procedure with respective change of variables $\alpha_{1,t} = \alpha$ and $k_t = \alpha\delta_{i,t}^2$ ∎

c) Hyperparameters. For ENET we will have:

$\frac{\partial \mathcal{L}}{\partial \alpha_{1,t}} = \frac{\partial}{\partial \alpha_{1,t}} \left\{ \frac{1}{2} log|\Sigma^{-1}(t)| + \frac{1}{2} log|diag(\Lambda_{.,t})| + \frac{1}{2} \mu_{.,t}^T \left(diag(\Lambda_{.,t})\right)^{-1} \mu_{.,t} \right\}$

Substituting $\Lambda_{i,t} = \bar{\Lambda}_{i,t}/\alpha_{1,t}$ and equating to zero:

$\frac{\partial \mathcal{L}}{\partial \alpha_{1,t}} = \sum_i \frac{\Sigma_{i,i}(t) + \mu_{i,t}^2}{\bar{\Lambda}_{i,t}} + \sum_i \frac{1}{2} \frac{\partial}{\partial \alpha_{1,t}} log \frac{1}{\alpha_{1,t}}$

$\alpha_{1,t} = (S/2)/\sum_i \left\{\frac{\Sigma_{i,i}(t) + \mu_{i,t}^2}{\bar{\Lambda}_{i,t}}\right\}$

For the hyperparameter defined by [2-5] we will have:

$\frac{\partial \mathcal{L}}{\partial k_t} = \frac{\partial}{\partial k_t} \left\{ -\sum_i logGa\left(\gamma_{i,t}/\frac{1}{2}, 1\right) + \sum_i log \int_{k_t}^\infty Ga\left(x/\frac{1}{2}, 1\right) dx - logp(k_t) \right\}$

Substituting $\gamma_{i,t} = k_t/(1-\bar{\Lambda}_{i,t})$:

$\frac{\partial \mathcal{L}}{\partial k_t} = \sum_i \frac{1}{1-\bar{\Lambda}_{i,t}} - S \frac{Ga(k_t/\frac{1}{2}, 1)}{\int_{k_t}^\infty Ga(x/\frac{1}{2}, 1) dx} + v - \left(\tau - \frac{S}{2}\right)\frac{1}{k_t}$

For the MXN model we have:

$\frac{\partial \mathcal{L}}{\partial \alpha} = \frac{\partial}{\partial \alpha} \sum_t \left\{ \frac{1}{2} log|\Sigma^{-1}(t)| + \frac{1}{2} log|diag(\Lambda_{.,t})| + \frac{1}{2} \mu_{.,t}^T \left(diag(\Lambda_{.,t})\right)^{-1} \mu_{.,t} - \sum_i logGa\left(\gamma_{i,t}/\frac{1}{2}, 1\right) + \sum_i log \int_{\alpha\delta_{i,t}^2}^\infty Ga\left(x/\frac{1}{2}, 1\right) dx - \frac{S}{2} log\alpha + \alpha\|W^{-1}\delta_{.,t}\|_1^2 \right\}$

$= \sum_{it} \frac{1}{2} \Sigma_{i,i}(t) \frac{\partial}{\partial \alpha} \frac{1}{\Lambda_{i,t}} + \sum_{it} \frac{1}{2} \frac{\partial}{\partial \alpha} log\Lambda_{i,t} + \sum_{it} \frac{1}{2} \mu_{i,t}^2 \frac{\partial}{\partial \alpha} \frac{1}{\Lambda_{i,t}} +$

$+ \sum_{it} \frac{1}{2} \frac{\partial}{\partial \alpha} log\alpha + \sum_{it} \frac{\delta_{i,t}^2}{1-\bar{\Lambda}_{i,t}} - \sum_{it} \frac{\delta_{i,t}^2 Ga(\alpha\delta_{i,t}^2/\frac{1}{2}, 1)}{\int_{\alpha\delta_{i,t}^2}^\infty Ga(x/\frac{1}{2}, 1) dx} -$

$- \frac{ST}{2} \frac{1}{\alpha} + \sum_t \|W^{-1}\delta_{.,t}\|_1^2$

Substituting $\Lambda_{i,t} = \bar{\Lambda}_{i,t}/\alpha$ and $\gamma_{i,t} = \alpha\delta_{i,t}^2/(1-\bar{\Lambda}_{i,t})$:

$\frac{\partial \mathcal{L}}{\partial \alpha} = \sum_{it} \frac{\Sigma_{i,i}(t) + \mu_{i,t}^2}{\bar{\Lambda}_{i,t}} + \sum_{it} \frac{\delta_{i,t}^2}{1-\bar{\Lambda}_{i,t}} + \sum_t \|W^{-1}\delta_{.,t}\|_1^2 -$

$- \frac{ST}{2} \frac{1}{\alpha} - \sum_{it} \frac{\delta_{i,t}^2 Ga(\alpha\delta_{i,t}^2/\frac{1}{2}, 1)}{\int_{\alpha\delta_{i,t}^2}^\infty Ga(x/\frac{1}{2}, 1) dx}$ ∎

d) Hyperparameters. In both ENET and MXN we will have:

$$\frac{\partial \mathcal{L}}{\partial \beta_t} = \frac{\partial}{\partial \beta_t}\frac{1}{2}log|\Sigma^{-1}(t)| + \frac{\partial}{\partial \beta_t}\frac{1}{2}log|\beta_t I|$$
$$+ \frac{\partial}{\partial \beta_t}\frac{1}{2\beta_t}\|V_{\cdot,t} - K\mu_{\cdot,t}\|_2^2$$
$$= \frac{\partial}{\partial \beta_t}\frac{1}{2}log|\Sigma^{-1}(t)| + \frac{N}{2}\frac{1}{\beta_t} - \frac{1}{2\beta_t^2}\|V_{\cdot,t} - K\mu_{\cdot,t}\|_2^2$$
$$= \frac{1}{2\beta_t}tr\left(\frac{\Sigma(t)}{diag(\Lambda_{\cdot,t})}\right) - S + \frac{N}{2}\frac{1}{\beta_t} - \frac{1}{2\beta_t^2}\|V_{\cdot,t} - K\mu_{\cdot,t}\|_2^2$$

Equating to zero we obtain:

$$\beta_t = \frac{\|V_{\cdot,t} - K\mu_{\cdot,t}\|_2^2}{N + \sum_{i=1}^S \left(\frac{\Sigma_{i,i}(t)}{\Lambda_{i,t}}\right) - S} \qquad \blacksquare$$

## D. Pseudo code for the algorithms

**Algorithm ENET-RVM**

INPUT: $K, V$
OUTPUT: $\mu, \gamma, \delta, \alpha$
For all $t = \overline{1, T}$.
    Initialize $\Lambda_{\cdot,t}$.
    Iterate until convergence criteria holds
        Compute $\Sigma(t)$ [2-43].
        Update $\mu_{\cdot,t}$ [2-31], $\overline{\Lambda}_{\cdot,t}$ [2-32], $\alpha_{1,t}$ [2-33], $k_t$ [2-34]
        and $\beta_t$ [2-36].
    End
End

**Algorithm MXN-RVM**

INPUT: $K, V$
OUTPUT: $\mu, \gamma, \delta, \alpha, \beta$
Initialize $\Lambda$.
Iterate until convergence criteria holds
    For all $t = \overline{1, T}$.
        Compute $\Sigma(t)$ [2-43].
        Update $\mu_{\cdot,t}$ [2-37], $\overline{\Lambda}_{\cdot,t}$ [2-38], $\delta_{\cdot,t}$ [2-39]
        and $\beta_t$ [2-40].
    End
Update $\alpha$ [2-32].
End